\documentclass{article}

\usepackage{arxiv}

\usepackage[utf8]{inputenc} 
\usepackage[T1]{fontenc}    
\usepackage{hyperref}       
\usepackage{url}            
\usepackage{booktabs}       
\usepackage{amsfonts}       
\usepackage{nicefrac}       
\usepackage{microtype}      
\usepackage{lipsum}
\usepackage{cite}
\usepackage{amsmath,amssymb,amsfonts}
\usepackage{algorithmic}
\usepackage{graphicx}
\usepackage{textcomp}
\usepackage{framed,multirow}
\usepackage[linesnumbered,ruled,lined]{algorithm2e}
\usepackage{makecell}
\usepackage{latexsym}
\usepackage{url}
\usepackage{float}
\usepackage[switch,pagewise,columnwise]{lineno}
\usepackage{marvosym}

\graphicspath{ {./images/} }

\title{Multiscale Attention Guided Network for COVID-19 Diagnosis Using Chest X-ray Images}

\author{
 Jingxiong Li \\
  Key Lab of RF Circuits and Systems of Ministry of Education\\Microelectronics CAD Center\\
  Hangzhou Dianzi University\\
  Hangzhou 310018, China\\
  \texttt{jingxiong.li2019@outlook.com} \\
   \And
 Yaqi Wang\textsuperscript{\Letter}\\
  Communication University of Zhejiang\\
  Hangzhou, China\\
  \texttt{wangyaqi@hdu.edu.cn} \\
  \And
 Shuai Wang \\
  Department of Radiology and BRIC\\
  University of North Carolina at Chapel Hill\\
  Chapel Hill NC 27599, USA\\
  \texttt{shuaiwang.tai@gmail.com} \\
  \And
 Jun Wang \\
  School of Biomedical Engineering\\
  Shanghai Jiao Tong University\\
  Shanghai, 200240, China\\
  \And
 Jun Liu \\
  Key Lab of RF Circuits and Systems of Ministry of Education\\Microelectronics CAD Center\\
  Hangzhou Dianzi University\\
  Hangzhou 310018, China\\
  \texttt{ljun77@hdu.edu.cn} \\
  \And
 Qun Jin \\
  School of Biomedical Engineering\\
  Shanghai Jiao Tong University\\
  Shanghai, 200240, China\\
  \texttt{jin@waseda.jp} \\
  \And
 Lingling Sun \\
  Key Lab of RF Circuits and Systems of Ministry of Education\\Microelectronics CAD Center\\
  Hangzhou Dianzi University\\
  Hangzhou 310018, China\\
  \texttt{sunll@hdu.edu.cn} \\

}

\begin{document}
\maketitle
\begin{abstract}
Coronavirus disease 2019 (COVID-19) is one of the most destructive pandemic after millennium, forcing the world to tackle a health crisis. Automated lung infections classification using chest X-ray (CXR) images could strengthen diagnostic capability when handling COVID-19.
However, classifying COVID-19 from pneumonia cases using CXR image is a difficult task because of shared spatial characteristics, high feature variation and contrast diversity between cases. Moreover, massive data collection is impractical for a newly emerged disease, which limited the performance of data thirsty deep learning models.
To address these challenges, Multiscale Attention Guided deep network with Soft Distance regularization (\textit{MAG-SD}) is proposed to automatically classify COVID-19 from pneumonia CXR images. In \textit{MAG-SD}, \textit{MA-Net} is used to produce prediction vector and attention from multiscale feature maps. To improve the robustness of trained model and relieve the shortage of training data, attention guided augmentations along with a soft distance regularization are posed, which aims at generating meaningful augmentations and reduce noise. Our multiscale attention model achieves better classification performance on our pneumonia CXR image dataset. Plentiful experiments are proposed for \textit{MAG-SD} which demonstrates its unique advantage in pneumonia classification over cutting-edge models. The code is available at \url{https://github.com/JasonLeeGHub/MAG-SD}.
\end{abstract}


\section{Introduction}
\label{sec:introduction}
The coronavirus disease 2019 (COVID-19) caused by severe acute respiratory syndrome coronavirus 2 (SARS-CoV-2) is one of the most devastating infectious diseases after millennium \cite{wang2020novel}. This new type of coronavirus is announced in late December, 2019, then spread globally in 2020. It has been declared as a pandemic by World Health Organization (WHO) according to its high contagiosity and unprecedented pressure bought to public healthcare system\cite{world2020coronavirus}. The current gold-standard for screening COVID-19 is polymerase chain reaction (PCR) laboratory test, however, the test capacity is extremely limited and requires professional equipment \cite{apostolopoulos2020extracting}. \cite{ai2020correlation} also reports that PCR tests suffers from high false negative rate.

Radiological images collected by X-ray and computed tomography (CT) are important complements to PCR tests. The virus leads to pneumonia, which is an inflammatory condition of the lung's air sacs \cite{huang2020clinical}. Radiological signs show ground-glass opacity, airspace opacities and later consolidation with bilateral, peripheral, and lower zone predominant distributions \cite{shi2020review}. Comparing with CT imaging, CXR diagnosis provides a low-cost and time-saving diagnosis method \cite{wang2019enhanced}. Besides, underdeveloped regions can hardly have sufficient CT scanners, making CT based COVID-19 screening impossible. X-rays are the most common diagnostic imaging equipment available even in rural regions, which means X-ray diagnosis can cover larger susceptible population \cite{franquet2001imaging}.

Diagnosis accuracy of COVID-19 and radiography based infection localization are critical for treatment planning and follow-up evaluations \cite{torres2015clinical}. However, pressure of pandemic forces physicians to evaluate in limited time, which raises misdiagnosis rate implicitly \cite{velavan2020covid}. As a result, accurate and robust classification methods are required. This is challenging as COVID-19 is a new type of disease which has low amount of data comparing with available datasets, such as image data published by \cite{bustos2019padchest} or \cite{irvin2019chexpert}. In addition, the COVID-19 shares characteristic with other types of pneumonia, which requires the method focus on both global and local features \cite{chen2020epidemiological}. Moreover, varied parameter settings causes imparities when collecting X-ray image from different devices.

Massive radiological data and rapid developing computational power give artificial intelligence (AI) a chance to assist clinical diagnosis. Recently, classification of COVID-19 from radiological images have been explored. Wang and Wong \cite{wang2020covid} present a COVID-Net operated on CXR images to classify COVID-19 from pneumonia and normal cases. COVID-19 cases are extracted from online COVID-19 datasets published by \cite{cohen2020covid} and \cite{COVID-chestxray-dataset}. Non-COVID-19 image includes 1591 pneumonia images and 1203 normal images released by National Institutes of Health Clinical Center \cite{wang2017chestx}.
The experimental results showed that classification method with residual projection-expansion-projection-extension (PEPX) design pattern achieves $93.3\%$ accuracy, which is better than general deep models such as VGG-19 ($83.0\%$ Accuracy) and ResNet-50 ($90.6\%$ Accuracy). The authors illustrate the locations focused by their model to visualize its decision making process.

Ghoshal and Tucker \cite{ghoshal2020estimating} present a Bayesian CNN to make diagnosis through model uncertainty. It is trained on 68 COVID-19 cases from \cite{cohen2020covid} and Non-COVID-19 cases from Kaggle’s Chest X-ray Images (Pneumonia) \cite{kermany2018identifying}, which improve the classification accuracy of a standard ResNet50V2 model from $86.0\%$ to $89.8\%$. The authors further discuss the effectiveness of uncertainty-aware classification by decision visualization.

Zhang et al. \cite{zhang2020covid} design a screening method based on ResNet to detect COVID-19 and find abnormalities from CXR images. Images are evaluated by an abnormity detecting module producing reference score to optimize classification loss. The model is trained on 70 COVID-19 images and 1008 non-COVID-19 images, which reaches $96.0\%$, $70.7\%$, $95.2\%$ in Sensitivity, Specificity and AUC respectively.

Generally, current studies operated on CXR images mostly depends on online datasets with limited COVID-19 cases. Insufficient data can hardly evaluate the robustness of the models and restricted their generalizability. Models trained on extremely imbalanced dataset also lead to long-tail distribution problems. Although plenty of works have discussed diagnosing COVID-19 by AI, few works address the problem of imbalanced data and limited size of dataset because of several issues: 1) Models trained by imbalanced data tend to classify all the targets to the dominant class which has overwhelmingly more labels than other classes. 2) Unique labels on X-ray image, such as L/R position labels, may easily attract model's attention then mislead the predictions. 3) COVID-19 cases share features with non-COVID cases, which requires a sensitive and robust model to do classification.

These challenges inspired us to treat pneumonia classification as a Fine-Grained Visual Classification (FGVC) problem, which aims at classifying sub-level categories under a basic-level category. FGVC cases are similar apart from some minor differences and also has the problem of lacking training data. Classic Convolutional Neural Networks (CNNs), including VGG \cite{simonyan2014very}, ResNet \cite{he2016deep} and Inception \cite{szegedy2016rethinking}, has difficulties handling this problem. We propose a novel Multiscale Attention Guided deep network with Soft Distance regularization (\textit{MAG-SD}) for COVID-19 CXR image classification. To balance the quantity of different data, a weakly-supervised method is presented, which requires a few labeled data to do effective augmentations. Multiscale strategy is applied to attention generator, producing detailed scalar matrix for prediction.
Our classification model is motivated from the fact that clinical diagnosis of COVID-19 follows a procedure which firstly evaluates the regional appearance, then makes diagnosis exclusively. Thus, we propose a multiscale attention module which estimates both shallow and deep layers. Comparing with using feature maps from only highest level features, the utilization of lower features could increase its ability of finding fine-grained features.
Moreover, a soft distance regularization method is integrated to refine classification result by adaptively adjusting classification loss. In a nutshell, contribution of this paper is threefold:

\textit{1)} We design a novel deep network, \textit{MA-Net}, to treat COVID-19 diagnosis as a FGVC problem. Multiscale attention is introduced to assess attention maps on multi level features. Composed attention maps are used as guidance for training steps. Attention pooling is proposed to utilize attention maps for classification.

\textit{2)} We address data shortage by proposing attention guided data augmentation and multi-shot training phase. It includes attention mix-up, attention patching and attention dimming that could enhance and search local feature then generating data. Models are trained on imbalanced COVID-19 datasets and achieve the state-of-the-art.

\textit{3)} Without introducing other modules or parameters, we formulate a new regularization term utilizing soft distance between predictions, which works as a constraint to limit classifiers from producing contradicted output for one target.

This paper is organized as follows. In Section. \ref{SEC:RLTDWRK}, we introduce insightful works which have high relevance with our contribution. Section. \ref{SEC:METHOD} presents the proposed method. In section. \ref{SEC:EXPERIMENT}, database and experimental setup are reported in detail, then results are presented and discussed individually. The last section concludes this study and highlights the future work.

\section{Related Works}
\label{SEC:RLTDWRK}
Related works are introduced in this section, including X-ray appearance for typical pneumonia, fine-grained visual classification, attention mechanism for CNNs and multiscale feature fusion utilized in computer vision.

\subsection{Pneumonia X-ray Imagery}
\label{SEC:PuXR}
Chest X-ray is a widely used imaging modality providing high-resolution pictures to visualize the pathological changes of thoracic diseases. Diagnosis could be made according to the visual patterns demonstrated on CXR images \cite{kermany2018identifying}. Clinical research from Katz and Leung \cite{katz1999radiology} demonstrated that typical image pattern for bacterial pneumonia includes opacity of single lobe and pleural effusion. Viral pneumonia also has radiological appearance such as pulmonary edema, small area of effusions, consolidation or lobe mass. Reports from \cite{rodriguez2020clinical}, \cite{rajaraman2020weakly}, demonstrated that the most common pattern on CXR in COVID-19 was consolidation or ground-glass opacity. It is notable that COVID-19 shares some visual feature with viral pneumonia while viral and bacterial pneumonia can hardly be differentiated because of similar spatial appearance.

\subsection{Fine-Grained Visual Classification}
Mass application of CNNs revealed its advantage in solving large scale image classification problem \cite{wei2019deep} and illuminated a promising way to settle FGVC tasks by using CNN models to explore inconspicuous local features. Some models relied on local annotations to train part-based detectors, localizing certain parts before prediction \cite{zhang2014part} \cite{wei2018mask}. However, local feature annotation requires expensive human labor, limiting its reproducibility in reality. Recently, approaches only require global labels also emerged whose motivation was to first localize the corresponding parts and then compare their local features \cite{lin2015bilinear}. Fu et al. \cite{fu2017look} introduced WS-DAN, which was a weakly supervised deep network handling FGVC by posing attention to enhance local feature and guide augmentation. FGVC was also a common problem in medical image because of spatial similarity between infections. Qin et al. \cite{qin2020fine} proposed a fine-grained classification CNN for different types of lung cancer in PET and CT Images.

\subsection{Attention for CNNs}
For visual task, attention usually indicates a scalar matrix representing the relative importance and inner relevance of local feature \cite{jetley2018learn}. This nonuniform representation was produced by special designed modules \cite{wang2020prior}. Works reported that applying attention on classification oriented CNN could provide an intuitional way to localize target object, helping to identify visual properties through local representation. An attention guided method demonstrated by Gondal et al. \cite{gondal2017weakly} reported that attention mechanism is helpful in Diabetic Retinopathy (DR) localization and recognition. Zhang et al. \cite{zhang2019attention} regulated the attention of deep model by training self-attention blocks for skin lesion classification and surpassed the baselines. Generally, attention mechanism guide the models to analyze global and local features simultaneously then generate believable classification results.

\subsection{Multiscale Feature Fusion}
Extracting hybrid feature maps from multi-resolution input image is a common strategy in computer vision since the the era of hand-engineered features. CNNs have an inherent multiscale feature in pyramidal shape, which is advantageous in producing semantically strong representations if effective feature fusion is operated. Models such as U-Net \cite{ronneberger2015u} and V-Net \cite{milletari2016v} exploited skip connections to associate feature maps across resolutions. FPN \cite{lin2017feature} leveraged the prediction of multiscale hierarchy by generating multiple prediction. For CXR image, Huang et al. \cite{huang2020fusion} presented weight concatenation method to cooperate global and local feature. Thriving of spatial attention gave inspiration to extract attention from multi-resolution feature map. Sedai et al. \cite{sedai2018deep} proposed A-CNN for chest pathologies localization, which utilized multiscale attention by calculating convex combination from weighted average of the feature maps.

\section{Method}
\label{SEC:METHOD}
In this episode, we propose our approach that explore multiscale fine-grain feature adaptively. We first produce an overview for our \textit{MAG-SD}. Then \textit{MA-Net} is presented in terms of network architecture with attention modules. A weakly supervised data augmentation module, \textit{Attention Guided Augmentation}, is introduced to address the shortage of COVID-19 cases. At last, \textit{Soft Distance Regularization} is proposed to erase noise imported by augmentations.
\subsection{Overview}
\begin{figure*}[!t]
\centering
\includegraphics[width=140mm]{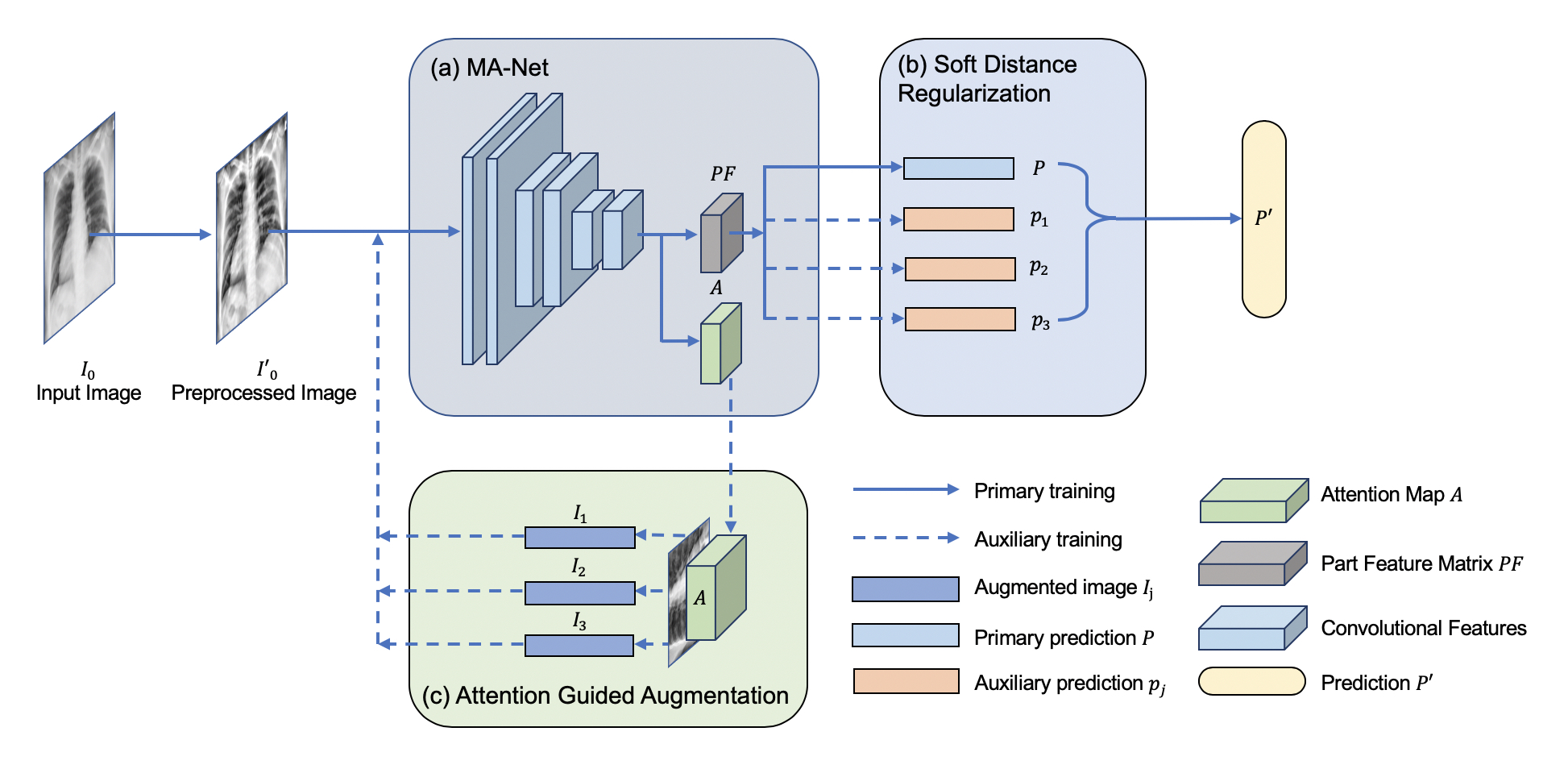}
\caption{
The architecture of \textit{MAG-SD}. The key components are illustrated in colour-wised blocks. (a): \textit{MA-Net}, which is a CNN model (\textit{e.g. ResNet50}) extracting prediction vectors $P, p_1, p_2, p_3$ and attention map $A$; (b): \textit{Soft Distance Regularization} using $P, p_1, p_2, p_3$ to calculate overall loss; (c): \textit{Attention Guided Augmentation}, which augments preprocessed data $I'_0$ according to $A$.}
\label{FigPPLine} 
\end{figure*}

COVID-19 CXR images are less distinctive comparing with other pneumonia cases, which requires a model to extract features for fine-grained feature of input image. WS-DAN\cite{fu2017look}, which is competitive in fine-grained image classification has been adopted for this topic. The architecture includes a feature extractor (\textit{i.e.} ResNet50), an attention generator operated on feature map and an augmentation generator producing local-enhanced and noise-blended image.
An overview of our \textit{MAG-SD} is shown in Fig. \ref{FigPPLine}. In primary training route, preprocessed CXR image $I'_0$ is fed into \textit{MA-Net} for prediction vector $P$ and attention map $A$. \textit{Attention Guided Augmentation} is operated on $I'_0$, using $A$ to produce augmented data $I_1, I_2, I_3$. In Auxiliary training routes, $I_1, I_2, I_3$ are pushed into \textit{MA-Net} for prediction vectors $p_1, p_2, p_3$. All the vectors (\textit{i.e. $P, p_1, p_2, p_3$}) are utilized by \textit{Soft Distance Regularization} for a proper loss.

\subsection{Multiscale Attention Guided Network (\textit{MA-Net})}
\subsubsection{Network Architecture}
\begin{figure}[!t]
\centering
\includegraphics[width=81mm]{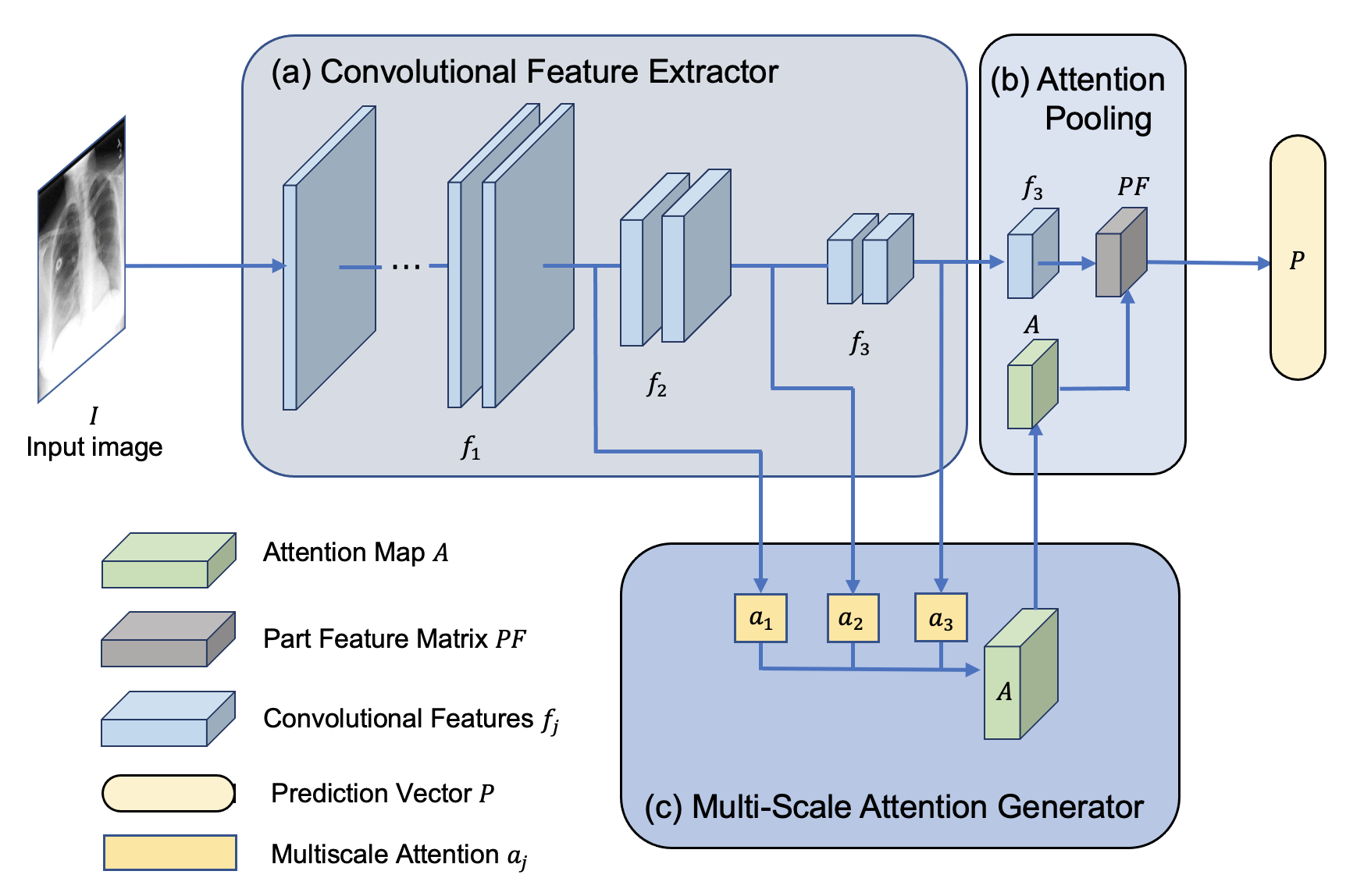}
\caption{\textit{MA-Net} illustrated in colour-wised blocks. (a): Convolutional Feature Extractor, which is a pretrained CNN model (e.g. ResNet50) extracting features $f_1, f_2, f_3$; (b): Attention Pooling (demonstrated in Fig. \ref{FigAttenP}) takes $f_3$ and attention map $A$ for prediction vector $P$); (c): Multiscale Attention Generator (demonstrated in Fig. \ref{FigAttenG}) uses $f_1, f_2, f_3$ to produce $A$ as output.}
\label{FigMANET} 
\end{figure}

Fig. \ref{FigMANET} presents a demonstration of our proposed \textit{MA-Net}. As observed, a CNN based encoder is operated on augmented images. Encoder utilizes ResNet50 as backbone, extracting size-different feature maps $f_1, f_2, f_3$ from image $I$. \textit{Multiscale attention generator} is used to extract attention map $a1, a2, a3$ and estimate scale-wised interests. Attention maps are resized for a single output $A$ from features. Then, the output of encoder $f_3$ and attention map $A$ are assessed by \textit{Attention Pooling} to generate prediction vector $P$.

\begin{figure}[!t]
\centering
\includegraphics[width=81mm]{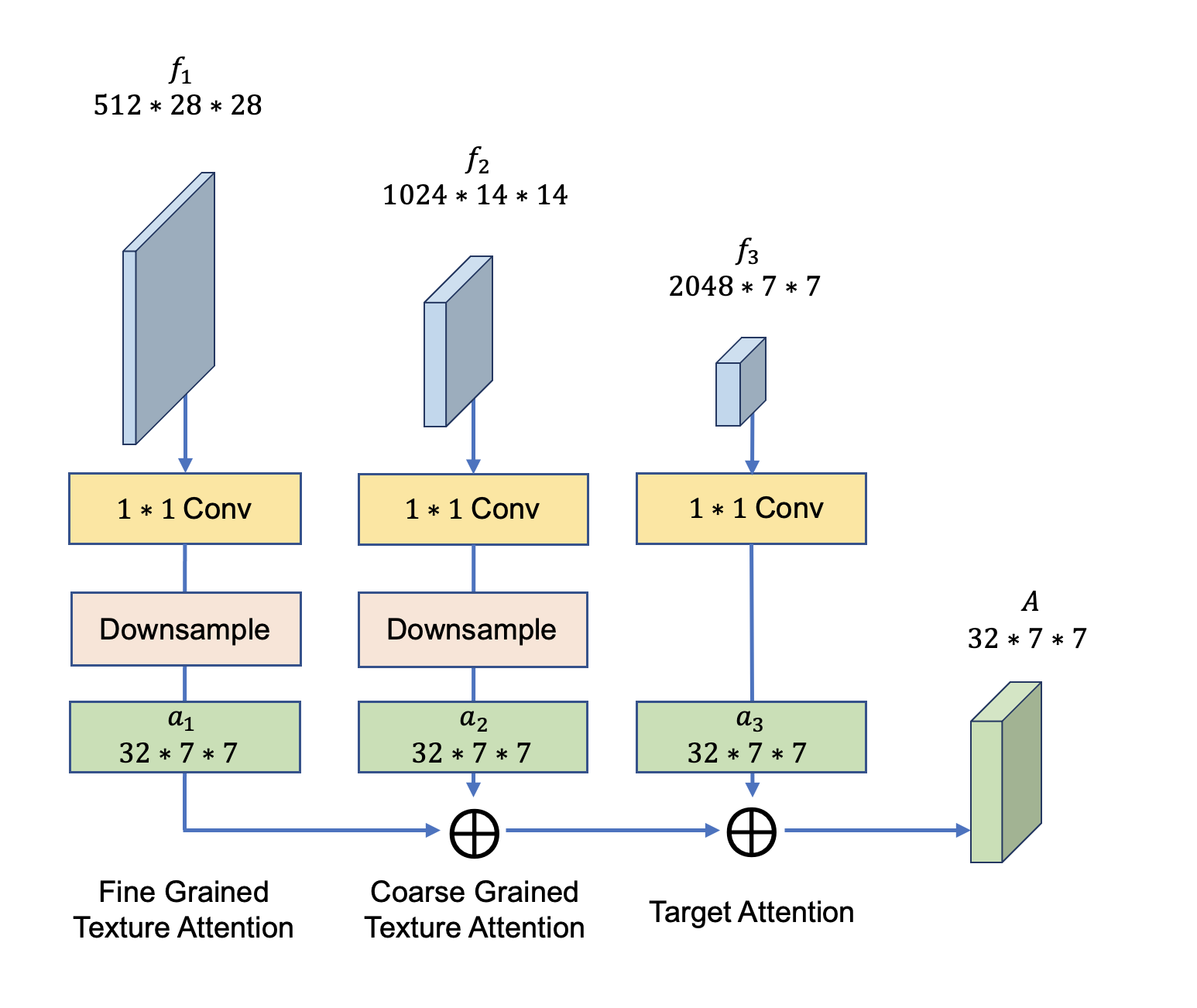}
\caption{Demonstration of multiscale attention generator. $f_1, f_2, f_3$ are three scales of feature maps. The model choose $1, 2$ or $3$ feature maps for attention. Attention map is generated by operating $1*1$ convolutional layer on each feature map then downsample it to $7*7$. Global attention map $A$ is produced by operating residual connection between resized feature maps. $\oplus$ represents residual connection.}
\label{FigAttenG} 
\end{figure}

\subsubsection{Multiscale Attention Generator}
Attention mechanism has been used in natural image topics to guide feedforward process \cite{jaderberg2015spatial}, \cite{chen2016attention}. Recently, tentative efforts have been made on deep models such as image classification \cite{wang2017residual}, person perception \cite{song2018mask} and sequential decision tasks \cite{noh2015learning}. Most of the attention models aim at gathering top level information to decide where to attend for the next learning steps. The proposed attention generating model is operated on multiscale feature maps, aiming at extracting attention from different scale. Layers before downsampling are selected as feature map in order to squeeze information out of single resolution feature. For ResNet50 we used, feature maps with $512*28*28, 1024*14*14, 2048*7*7$ sizes are chosen. The number of attention map is $32$.

The architecture of multiscale attention generator is shown in Fig. \ref{FigAttenG}. $f_1, f_2$ and $f_3$ are feature maps selected from feature extractor. Each of them are processed by $1*1$ convolution to generate corresponding attention. All the attention maps are downsampled to $7*7$ and connected residually. The effect of using different number of feature maps is discussed in experiments.

\begin{figure}[!t]
\centering
\includegraphics[width=81mm]{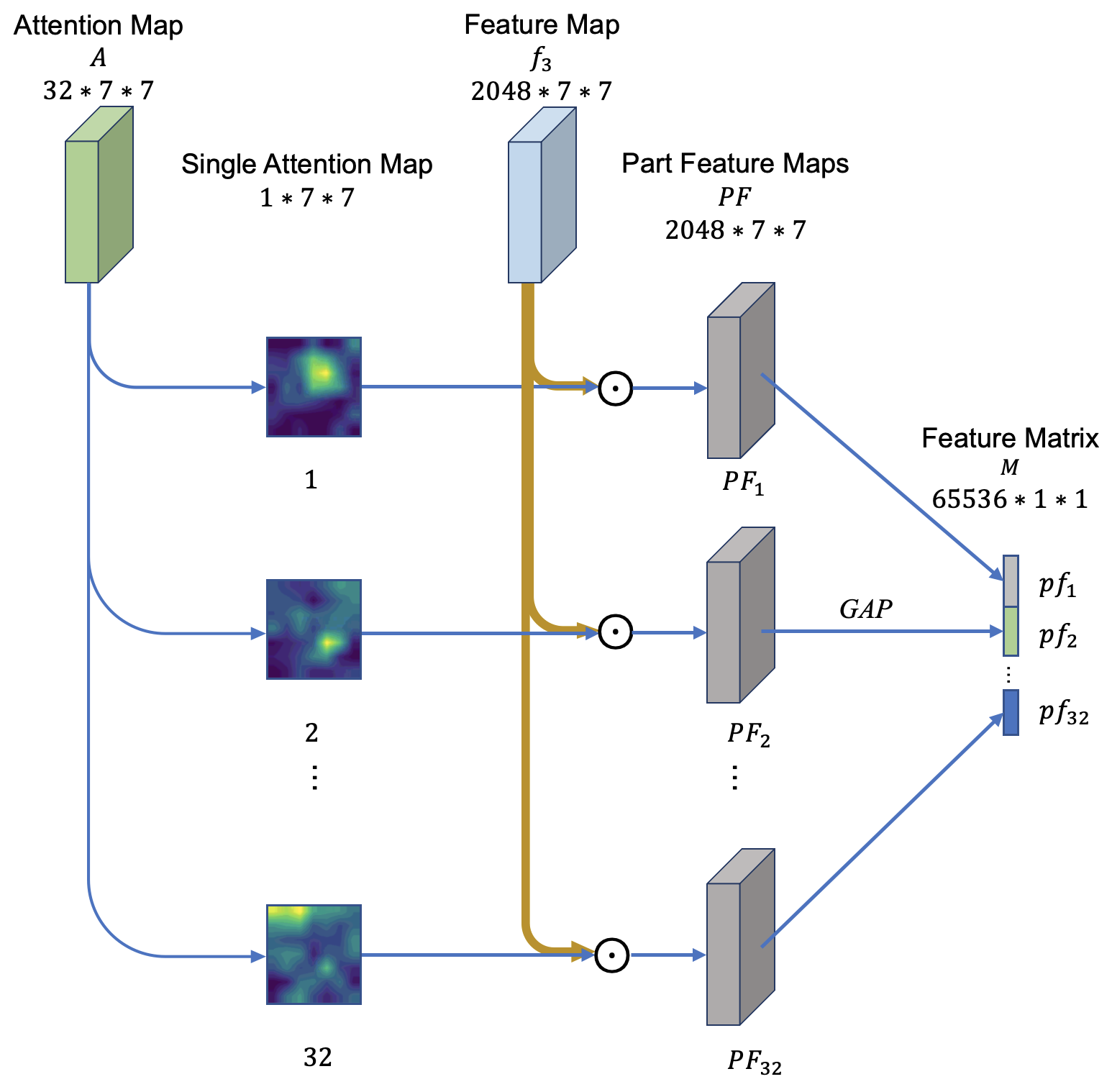}
\caption{Attention pooling architecture proposed for feature selecting. Feature map $f_3$ is extracted from input image and $A$ is generated by the module displayed in Fig. 3. Each individual attention map selected from $A$ multiplies ($\odot$) with $f_3$ to produce the features with attention bias, known as part feature Maps ($PF$). After global average pooling (GAP) process, feature matrix $M$ is produced.}
\label{FigAttenP} 
\end{figure}

\subsubsection{Attention Pooling}
Attention pooling module mimics the structure proposed by \cite{fu2017look}, which associates attention output and feature map. Fig. \ref{FigAttenP} shows the pipeline of the pooling method. Feature map $f_3$ ($2048*7*7$) is extracted from the output of CNN encoder. Multiscale attention map $A$ presented by attention generator is $32*7*7$. Each attention map focuses on diverse location that may contain valuable fine-grained feature. Attention biased features (\textit{i.e. part feature map ($PF$)}) are presented by multiplying all the attention maps $A$, each by each, with feature map. There are $32$ $PF$s which size equals $2048*7*7$. Global average pooling (GAP) is operated to shrink each $PF$ to $2048*1*1$ in order to describe the activation intensity of attention on feature map. Feature matrix $M$ is produced by concatenating GAP results, producing a vector of $65536*1*1$. Eq. (\ref{Eq2}) describes the calculation of $PF$.
\begin{equation}
\label{Eq2}
    PF_j = A_j\odot f_3 \;\; (j=1,2,...,N)
\end{equation}
where $\odot$ stands for multiplication of elements between two tensors. $f_3$ is feature map extracted by CNN. $N$ represents the number of attention maps, which is 32 in our work.

$PF_j$ has to go through a downsampling method such as GAP to get description with compressed size, which is $2048*1*1$.
Feature matrix $M$ is represented by concatenating all condensed $PF_j$ presented in Fig. \ref{FigAttenP}.

\begin{figure}[!t]
\centering
\includegraphics[width=81mm]{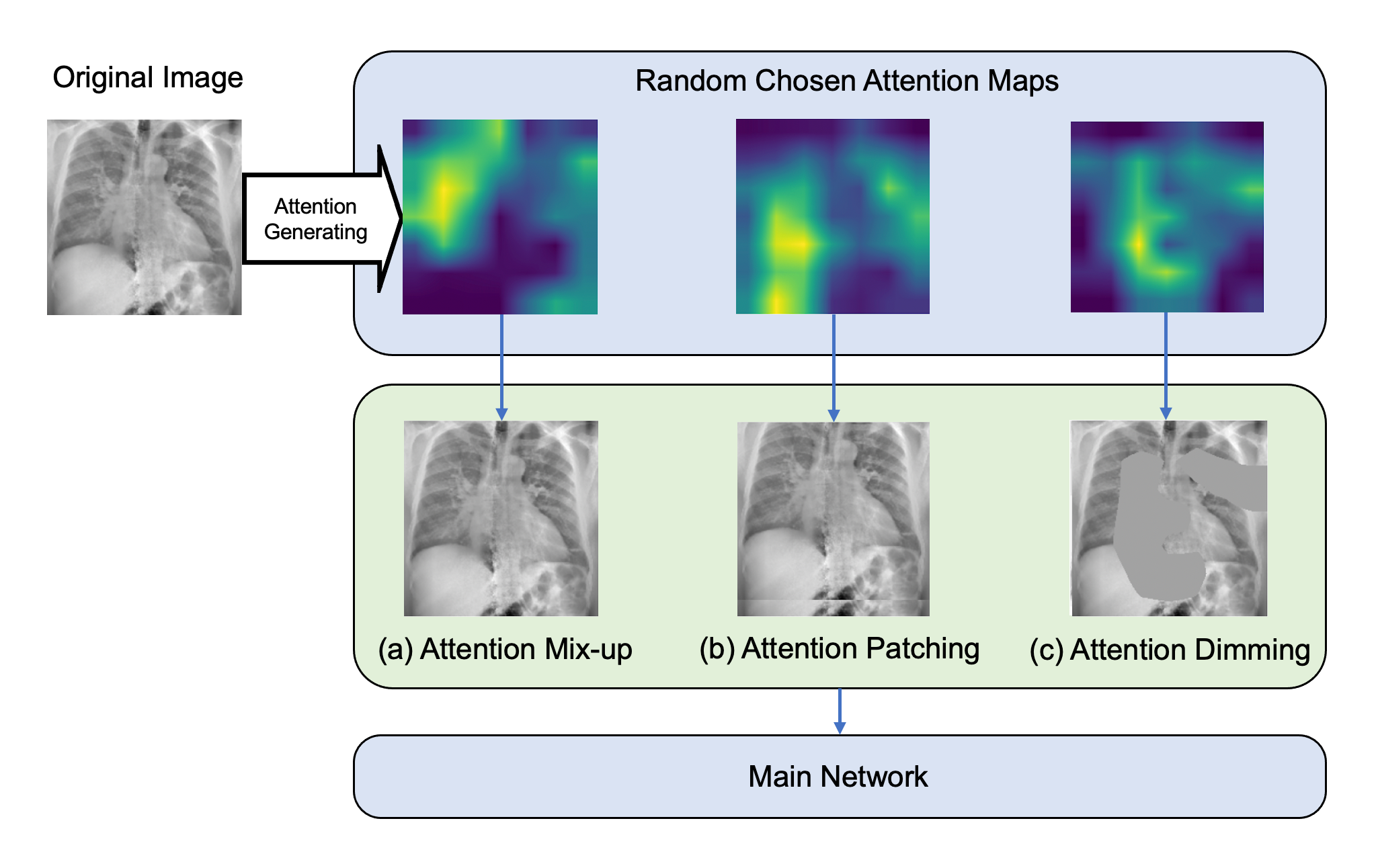}
\caption{Demonstration of Attention Guided Augmentation. Multiple attention maps are generated by attention generator, which concentrate on different part of original image. One attention map is chosen randomly for each augmentation method, including: (a)Attention mix-up, (b)Attention Patching and (c)Attention Dimming.}
\label{FigReT}
\end{figure}

\subsection{Attention Guided Augmentation}
As mentioned above, attention mechanism emphasizes local feature which affects the classification result. Following the idea, the performance of classification network could be enhanced if attention guided training cases are considered. Weakly supervised methods demonstrated in Fig. \ref{FigReT} present effective augmentations for original image. One normalized attention map ($A^*$) is randomly chosen for each instance to do individual augmentation.
\textit{1) Attention Mixup: }Mixup is an augmentation strategy which generates data by mixing overall image and regional feature together. As we have a attention map $A^*_j$, a detailed region $D_j$could be extracted by doing threshold.
\begin{equation}
\label{eq6}
    D(l,m) = \begin{cases}
    1,&\text{if}\; A^*(l,m)>\theta_m \\
    0,&\text{otherwise}
    \end{cases}
\end{equation}

For elements in $ A_j^*(l,m)$, Eq. (\ref{eq6}) sets $D_j(l,m)$ to $1$ if it is greater than threshold $\theta_m \in [0,1]$. If not, it will be set to $0$. A bounding box surrounding the extracted region is proposed from the raw region. Region coved by the box is enlarged to the same size as input image then merged together with original input $I_0$ to get augmented input $I_1$, which is defined in Eq. (\ref{eq7}).
\begin{equation}
\label{eq7}
    I_1(p,q) = \gamma I_0(p,q) + (1-\gamma) B(p,q)
\end{equation}
where $\gamma$ is a parameter range in $[0,1]$ and $B$ stands for the enlarged bounding box. Model could see target precisely by learning local and global feature together.

\textit{2) Attention Patching: }Encoder could be sensitive to limited part of reception field as valuable spatial features usually distribute in similar position. To encourage the encoder to explore feature from varied part, attention patching is proposed. $D$ mentioned in \textit{1)} is patched onto the original image $I_0$ to propose patched data $I_2$, which is demonstrated in Fig. \ref{FigReT}.
Attention patching enlarges the model's interest region by duplicating the interested area, promoting model to global evaluate its input.

\textit{3) Attention Dimming: }When training attention generating module for feature map, multiple attention maps may be sensitive to similar region. A responsible fine-grain classification model have to focus on different local features of one target. Attention dimming is proposed to stimulate the attention model searching the whole reception field for valuable information. We obtain a Dimming Mask ($DM$) from $ A^*$, applying threshold $\theta_d\in[0,1]$, as represented in Eq. (\ref{eq8}).
\begin{equation}
\label{eq8}
    DM(l,m) = \begin{cases}
    00.1,&\text{if}\; A^*(l,m)>\theta_d \\
    1,&\text{otherwise}
    \end{cases}
\end{equation}

Augmented image $I_3$ is generated by applying the mask onto the input, which is illustrated in Fig. \ref{FigReT}(c)

\subsection{Soft Distance Regularization}
Disturbances are introduced into the original image by using augmentation. (\textit{e.g. infection area reduced by attention dimming}).
To address this problem, we formulate the uncertainty of predictions via the distance between prediction vectors. Intuitively, the distance $d$ could be modeled as Eq. (\ref{eq10}).
\begin{equation}
\label{eq10}
     d(x) = \left | P(I) - p(x)\right |
\end{equation}
where $x$ denotes the augmented image, $P(I)$, $p(x)$ represent primary and auxiliary prediction vector respectively. However, the distance between $P(I)$ and $p(x)$ is unstable before the the model well-fitted. Ground truth labels are referenced to stabilize gradients. In Algorithm \ref{Alg1}, $P(I)$ is replaced by soft label $P'(I)$, filtering out low confidence inferences. Soft distance $d'(x)$ can be represented in Eq. (\ref{eq11}). The value of $\theta$ in Algorithm. \ref{Alg1} is $0.7$.
\begin{equation}
\label{eq11}
     d'(x) = \left | P'(I) - p(x)\right |
\end{equation}

\begin{algorithm}
\caption{Soft Distance Regularization}
\label{Alg1}
\LinesNumbered
\KwIn{
$P(I)$: Primary prediction vector\newline
$p(x)$: Auxiliary prediction vector\newline
$G_{lbl}(I)$: Ground truth labels\newline
$\theta$: Confidence threshold
}
\KwOut{$L_{reg}$: soft distance regularization term}

Cross entropy loss $L_{ce}^{prim}$ is calculated between $P(I)$ and $G_{lbl}(I)$\;
$P(I)$, $p(x)$ are fed into \textit{softmax} to extract confidence score over all classes, which are $P^c(I)$, $p^c(x)$\;
\uIf{$P^c(I) > \theta$}{Let $P'(I) = P(I)$}
\Else{Let $P'(I) = G_{lbl}(I)$}
Predict variance is represented by soft distance between $P'(I)$ and $p(x)$:
$$ d'(x) = \left | P'(I) - p(x)\right |$$\\
Overall loss is combined by $L_{ce}^{prim}$ and mean predicting variance:
$$ L_{reg}= L_{ce}^{prim} + \bar{d'}$$\\
\textbf{return} $L_{reg}$
\end{algorithm}

At last, overall loss is modeled by a combination of cross entropy loss and average soft distance, which is demonstrated in Eq. (\ref{eq12}).
\begin{equation}
\label{eq12}
     L_{reg}= L_{ce}^{prim} + \bar{d'}
\end{equation}

where $L_{ce}^{prim}$ operates between labels and primary prediction. If two vectors have different prediction for one target, $L_{reg}$ will generate a large value, which reflects the uncertainty of the model on one target. It is also notable that $L_{reg}$ punishes soft distance $\bar{d'}$, leading the model to generate similar predictions.

\section{Experiments}
\label{SEC:EXPERIMENT}

In this section, extensive experiments were conducted to comprehensively assess \textit{MAG-SD}. Models were trained on datasets with different types of pneumonia and the performance of each proposed method was evaluated. Then the models were compared between other baseline methods using several metrics.

\subsection{Dataset and Experimental Settings}
The proposed model was trained and tested on several datasets to evaluate its classification performance and ability of fine-grained pneumonia localization. Details of each dataset was shown in Tab. \ref{TAB:Sets}. \textit{Dataset A} was a mutated dataset with 90 COVID-19 from \cite{cohen2020covid} and 168 other pneumonia cases from \cite{wang2017chestx}, which directly assessed model's fine-grained classification ability. \textit{Dataset B} was selected from \cite{LungSeg_dataset} and \cite{wang2017chestx}, aiming at evaluating the model's performance on larger scale. \textit{Dataset C} was the largest dataset we operated on, which included COVID-19 detection and fine-grained pneumonia classification. Quality of pneumonia localization was evaluated by \textit{Localization} dataset, which had $13$ COVID-19 cases with pixel-wise masks from \cite{vaya2020bimcv} and $118$ non-COVID pneumonia cases with bounding boxes annotations from \cite{wang2017chestx}.
In experiments, classic ResNet50 has been adopted as feature extractor. Its \textit{layer4} output was chosen as feature map. Attention was extracted from the output of \textit{layer2, layer3} and \textit{layer4} to ensure multiscale attention. Size of the attention maps were $28*28,14*14$ and $7*7$. Both training and testing sets were divided roughly in the same class proportions. 5-fold cross validation was applied to get reliable results.

Models were implemented using Pytorch and trained on two NVIDIA RTX 2080TI GPUs. The optimizer was Stochastic Gradient Descent (SGD) with the momentum of 0.9. For each training, 100 training epochs were deployed, with $10^{-6}$ weight decay, $32$ cases per minibatch and $10^{-3}$ learning rate at beginning. Images were resized to $224*224$ when training and testing.

\begin{table}[!t]
\renewcommand{\arraystretch}{1.1}
\caption{Datasets details}
\label{TAB:Sets}
\centering
\begin{tabular}{cccc}
\Xhline{1.2pt}
\textbf{Dataset}           & \textbf{Class}         & \textbf{Value} & \textbf{Total}        \\ \hline
\multirow{2}{*}{Dataset A} & COVID-19               & 90             & \multirow{2}{*}{258}  \\
                           & Non-COVID-19 pneumonia & 168            &                       \\ \hline
\multirow{3}{*}{Dataset B} & COVID-19               & 462            & \multirow{3}{*}{3631} \\
                           & Non-COVID-19 pneumonia & 1567           &                       \\
                           & Healthy                & 1602           &                       \\ \hline
\multirow{4}{*}{Dataset C} & COVID-19               & 462            & \multirow{4}{*}{6329} \\
                           & Viral pneumonia        & 1449           &                       \\
                           & Bacterial pneumonia    & 2816           &                       \\
                           & Healthy                & 1602           &                       \\ \hline
\multirow{2}{*}{Localization}& COVID-19               & 13             & \multirow{2}{*}{131}  \\
                           & Non-COVID-19 pneumonia & 118            &                       \\ \Xhline{1.2pt}
\end{tabular}
\end{table}

\subsection{Pre-Processing and Data Augmentation}

X-ray images have different appearance according to varied imaging equipment configurations, resolving that the same tissue can be radiologically different. To ensure the intensity distribution of one tissue is similar over the dataset, Z-score normalization was employed when training and testing. Large contrast distribution also introduced extra noise to the dataset, impacting the performance of trained model. Contrast limited adaptive histogram equalization (CLAHE) was proposed to enhance contrast between tissues and restrain noise signal \cite{pisano1998contrast}.

In image classification, data augmentation has been proved as an effective method to improve robustness and evaluate performance \cite{perez2017effectiveness}. Augmented data provides more varieties for classification target and remitting the impact of overfitting. Random number of transformations were chosen from a sequence of linear transformation for each training sequence. The list is shown in Tab. \ref{TAB:Augs}.

\begin{table}[!t]
\renewcommand{\arraystretch}{1.1}
\caption{Augmentations used and factor setting}
\label{TAB:Augs}
\centering
\begin{tabular}{ll}
\Xhline{1.2pt}
\multicolumn{1}{c}{\textbf{Augmentations}} & \multicolumn{1}{c}{\textbf{Abstract}}   \\ \hline
\multicolumn{1}{c}{Brightness adjustment}& \begin{tabular}{c}Random chosen brightness \\factor from $[0.5,1.0]$\end{tabular} \\\hline
\multicolumn{1}{c}{Contrast adjustment} & \begin{tabular}{c}Random chosen contrast \\factor from $[0.7,1.0]$\end{tabular}   \\\hline
\multicolumn{1}{c}{Resized cropping}    & \begin{tabular}{c}Random cropping then \\resize to $224*224$\end{tabular}  \\\hline
\multicolumn{1}{c}{Rotation}            & Random rotation from $[0,120]$            \\\hline
\multicolumn{1}{c}{Horizontal flipping} & \multicolumn{1}{c}{-}                     \\\hline
\multicolumn{1}{c}{Vertical flipping}   & \multicolumn{1}{c}{-}                     \\\Xhline{1.2pt}
\end{tabular}
\end{table}

\begin{table*}[h]
\caption{Evaluation of Models}
\centering
\label{TAB:Models}
\renewcommand{\arraystretch}{1.1}
\resizebox{\textwidth}{!}{
\begin{tabular}{c|cccc|cccc|cccc}
\Xhline{1.2pt}
\multirow{2}{*}{\textbf{Model}} & \multicolumn{4}{c|}{\textbf{Dataset A}}          & \multicolumn{4}{c|}{\textbf{Dataset B}}                  & \multicolumn{4}{c}{\textbf{Dataset C}} \\ \cline{2-13}
                       &ACC($\%$)           & SEN($\%$)   &SPC($\%$)   & F1($\%$)    & ACC($\%$)   & SEN($\%$)            &SPC($\%$)            & F1($\%$)    & ACC($\%$)    & SEN($\%$)   &SPC($\%$)   & F1($\%$)    \\ \hline
VGG16\cite{simonyan2014very} & 92.88$\pm$1.35 & 91.51$\pm$1.17 & 92.77$\pm$1.89 & 92.08$\pm$1.44 & 90.68$\pm$1.64 & 91.05$\pm$2.12    & 94.90$\pm$0.82   & 89.44$\pm$1.98 & 80.23$\pm$1.44  & 77.97$\pm$2.02 & 92.98$\pm$0.43 & 78.82$\pm$1.28 \\
ResNet18\cite{he2016deep}& 90.94$\pm$1.39 & 91.87$\pm$1.55 & 88.56$\pm$1.79 & 89.83$\pm$1.46 & 92.06$\pm$1.14 & 92.03$\pm$1.13          & 95.62$\pm$1.03          & 91.12$\pm$1.34 & 82.33$\pm$1.35  & 82.61$\pm$1.20 & 93.57$\pm$1.22 & 81.30$\pm$1.23 \\
ResNet50\cite{he2016deep}& 92.94$\pm$1.19 & 91.16$\pm$1.14 & 94.25$\pm$1.53 & 92.31$\pm$1.39 & 92.56$\pm$0.76   & 92.07$\pm$1.68     & 95.85$\pm$0.47   & 91.74$\pm$0.93 & 82.94$\pm$1.02  & 84.01$\pm$1.16 & 93.61$\pm$0.37 & 82.64$\pm$1.39 \\
InceptionV3\cite{szegedy2016rethinking}& 94.13$\pm$1.13 & 92.94$\pm$1.02 & 94.49$\pm$1.11 & 93.62$\pm$1.12 & 93.06$\pm$1.19 & 91.73$\pm$1.13      & 96.23$\pm$0.77    & 92.42$\pm$1.51 & 84.20$\pm$1.19  & 85.19$\pm$1.35 & 94.03$\pm$0.99 & 84.69$\pm$1.27 \\ \hline
\cite{narin2020automatic}(ResNet) & 93.88$\pm$1.18 & 93.01$\pm$1.32 & 93.63$\pm$1.36 & 93.30$\pm$1.55 & 93.41$\pm$1.14 & 93.71$\pm$1.72       & 96.26$\pm$0.62  & 93.12$\pm$1.49 & 83.93$\pm$1.11  & 86.40$\pm$0.99 & 93.85$\pm$0.42 & 84.39$\pm$1.01 \\
\cite{narin2020automatic}(InceptionV3) & 94.56$\pm$1.75     & 92.19$\pm$1.62 & 94.91$\pm$1.47 & 93.40$\pm$1.47 & 93.45$\pm$1.21 & 93.00$\pm$1.24    & 96.40$\pm$0.50  & 93.04$\pm$1.02 & 84.93$\pm$1.67  & 85.90$\pm$1.01 & 94.36$\pm$0.71 & 84.92$\pm$1.70 \\
COVID-Net\cite{wang2020covid} & 93.25$\pm$1.70 & 91.62$\pm$1.88 & 93.70$\pm$1.80 & 92.51$\pm$1.89 & 88.94$\pm$1.28 & 89.95$\pm$2.41          & 93.75$\pm$0.58          & 87.98$\pm$1.56 &78.71$\pm$1.76   & 79.26$\pm$0.94 & 92.38$\pm$0.47 & 78.34$\pm$1.40 \\ \hline
BCNN\cite{lin2015bilinear} & 96.00$\pm$1.52 & 96.43$\pm$1.78 & 94.52$\pm$1.60 & 95.39$\pm$1.42 & 94.41$\pm$1.37 & 95.26$\pm$1.23 & 96.71$\pm$0.81 & \textbf{96.71$\pm$1.60} & 84.36$\pm$1.84  & 84.47$\pm$0.75 & 94.15$\pm$0.49 & 84.47$\pm$0.85 \\
BCNN(Attention(Ours))        & 96.43$\pm$1.45 & 96.31$\pm$1.55 & \textbf{96.16$\pm$1.74} & 96.23$\pm$1.30 & 95.11$\pm$1.55 & \textbf{96.61$\pm$2.00}   & 97.26$\pm$0.72  & 94.12$\pm$1.92 & 85.04$\pm$2.36  & 86.32$\pm$1.55 & 94.33$\pm$0.86 & 84.41$\pm$1.52 \\ \hline
FPN\cite{lin2017feature} & 94.88$\pm$1.61 & 95.11$\pm$1.75 & 94.30$\pm$1.22 & 94.65$\pm$1.72 & 93.27$\pm$1.20    & 94.05$\pm$0.82    & 96.20$\pm$0.78   & 92.86$\pm$1.11 & 82.17$\pm$1.89  & 83.58$\pm$1.42 & 93.32$\pm$0.50 & 81.85$\pm$1.16 \\
U-Net\cite{ronneberger2015u}& 95.76$\pm$1.07 & 94.92$\pm$1.10 & 95.91$\pm$1.62 & 95.38$\pm$1.13 & 93.00$\pm$1.55 & 92.59$\pm$1.39 & 96.08$\pm$0.82    & 92.33$\pm$1.84 & 84.01$\pm$1.36  & 84.68$\pm$1.42 & 94.14$\pm$0.56 & 83.81$\pm$1.24 \\ \hline
MAG-SD(0AUG) & 94.31$\pm$1.28  & 91.70$\pm$1.33  & 95.89$\pm$1.94 & 93.37$\pm$1.38 & 93.25$\pm$0.90 & 92.01$\pm$0.93  & 96.29$\pm$0.45          & 92.06$\pm$1.17 & 84.13$\pm$0.99  & 85.08$\pm$1.27 & 94.11$\pm$0.43 & 84.30$\pm$1.43 \\
\textbf{MAG-SD(Proposed)} &
  \textbf{96.94$\pm$1.10} &
  \textbf{97.83$\pm$1.53} &
  94.93$\pm$1.26 &
  \textbf{96.23$\pm$1.02} &

  \textbf{95.85$\pm$1.27} &
  95.74$\pm$1.20 &
  \textbf{97.73$\pm$0.45} &
  95.54$\pm$1.59 &
  \textbf{87.12$\pm$1.55} &
  \textbf{87.20$\pm$1.64} &
  \textbf{95.20$\pm$0.64} &
  \textbf{86.98$\pm$1.27} \\ \Xhline{1.2pt}
\end{tabular}}
\end{table*}

\begin{table*}[]
\renewcommand{\arraystretch}{1.1}
\caption{Evaluation of CLAHE}
\label{TAB:CLAHE}
\centering
\resizebox{\textwidth}{!}{
\begin{tabular}{c|cccc|cccc|cccc}
\Xhline{1.2pt}
\multirow{2}{*}{\textbf{Preprocessing}} & \multicolumn{4}{c|}{\textbf{Dataset A}} & \multicolumn{4}{c|}{\textbf{Dataset B}} & \multicolumn{4}{c}{\textbf{Dataset C}} \\ \cline{2-13}
                       &ACC($\%$)           & SEN($\%$)   &SPC($\%$)   & F1($\%$)    & ACC($\%$)   & SEN($\%$)            &SPC($\%$)            & F1($\%$)    & ACC($\%$)    & SEN($\%$)   &SPC($\%$)   & F1($\%$)   \\ \hline
w/o CLAHE & 95.56$\pm$1.14 & 93.12$\pm$1.59 & 96.23$\pm$1.03 & 94.50$\pm$0.90 & 93.45$\pm$1.58 & 92.75$\pm$2.17 & 96.37$\pm$1.40 & 92.64$\pm$1.23 & 85.47$\pm$1.20  & 86.64$\pm$1.96 & 94.59$\pm$0.85 & 85.96$\pm$1.64 \\
CLAHE&
  \textbf{96.94$\pm$1.10} &
  \textbf{97.83$\pm$1.53} &
  \textbf{94.93$\pm$1.26} &
  \textbf{96.23$\pm$1.02} &

  \textbf{95.85$\pm$1.27} &
  \textbf{95.74$\pm$1.20} &
  \textbf{97.73$\pm$0.45} &
  \textbf{95.54$\pm$1.59} &
  \textbf{87.12$\pm$1.55} &
  \textbf{87.20$\pm$1.64} &
  \textbf{95.20$\pm$0.64} &
  \textbf{86.98$\pm$1.27} \\ \Xhline{1.2pt}
\end{tabular}}
\end{table*}

\begin{table*}[]
\renewcommand{\arraystretch}{1.1}
\caption{Evaluation of Multisize Attention Maps}
\label{TAB:FeatureSizes}
\centering
\resizebox{\textwidth}{!}{
\begin{tabular}{ccccccccccccc}
\Xhline{1.2pt}
\multicolumn{1}{c|}{\multirow{2}{*}{\textbf{Attention Maps}}} &
  \multicolumn{4}{c|}{\textbf{Dataset A}} &
  \multicolumn{4}{c|}{\textbf{Dataset B}} &
  \multicolumn{4}{c}{\textbf{Dataset C}} \\ \cline{2-13}
\multicolumn{1}{c|}{} &
  ACC($\%$)&
  SEN($\%$)&
 SPC($\%$)&
  \multicolumn{1}{c|}{F1($\%$)} &
  ACC($\%$)&
  SEN($\%$)&
 SPC($\%$)&
  \multicolumn{1}{c|}{F1($\%$)} &
  ACC($\%$)&
  SEN($\%$)&
 SPC($\%$)&
  F1($\%$)\\ \hline
\multicolumn{1}{c|}{7*7} &
  95.31$\pm$1.10 &
  96.70$\pm$1.49 &
  91.09$\pm$1.04 &
  \multicolumn{1}{c|}{93.48$\pm$1.07} &
  94.75$\pm$1.54 &
  94.93$\pm$1.44 &
  96.98$\pm$1.06 &
  \multicolumn{1}{c|}{94.11$\pm$1.57 } &
  85.44$\pm$1.47 &
  86.56$\pm$0.96 &
  94.56$\pm$0.36 &
  86.36$\pm$0.92\\
\multicolumn{1}{c|}{7*7 + 14*14} &
  \textbf{96.94$\pm$1.10} &
  \textbf{97.83$\pm$1.53} &
  \textbf{94.93$\pm$1.26} &
  \multicolumn{1}{c|}{\textbf{96.23$\pm$1.02}} &
  \textbf{95.85$\pm$1.27}&
  \textbf{95.74$\pm$1.20} &
   \textbf{97.73$\pm$0.45}&
  \multicolumn{1}{c|}{\textbf{95.54$\pm$1.59}} &
  \textbf{87.12$\pm$1.55} &
  \textbf{87.20$\pm$1.64} &
  \textbf{95.20$\pm$0.64} &
  \textbf{86.98$\pm$1.27} \\
\multicolumn{1}{c|}{7*7 + 14*14 + 28*28} &
  \multicolumn{1}{c}{95.88$\pm$1.84} &
  \multicolumn{1}{c}{94.76$\pm$2.11} &
  \multicolumn{1}{c}{96.63$\pm$2.23} &
  \multicolumn{1}{c|}{95.54$\pm$1.91} &
  \multicolumn{1}{c}{94.74$\pm$1.79} &
  \multicolumn{1}{c}{95.80$\pm$2.49} &
  \multicolumn{1}{c}{96.92$\pm$1.17} &
  \multicolumn{1}{c|}{94.57$\pm$2.16} &
  \multicolumn{1}{c}{85.01$\pm$1.89} &
  \multicolumn{1}{c}{86.65$\pm$2.29} &
  \multicolumn{1}{c}{94.29$\pm$0.97} &
  \multicolumn{1}{c}{85.73$\pm$2.25} \\ \Xhline{1.2pt}
\end{tabular}}
\end{table*}

\begin{table*}[]
\renewcommand{\arraystretch}{1.1}
\caption{Comparison of Pooling Methods}
\label{TAB:Poolings}
\centering
\resizebox{\textwidth}{!}{
\begin{tabular}{c|cccc|cccc|cccc}
\Xhline{1.2pt}
\multirow{2}{*}{\textbf{Pooling}} & \multicolumn{4}{c|}{\textbf{Dataset A}}          & \multicolumn{4}{c|}{\textbf{Dataset B}} & \multicolumn{4}{c}{\textbf{Dataset C}} \\ \cline{2-13}
    &ACC($\%$)           & SEN($\%$)   &SPC($\%$)   & F1($\%$)    & ACC($\%$)   & SEN($\%$)            &SPC($\%$)            & F1($\%$)    & ACC($\%$)    & SEN($\%$)   &SPC($\%$)   & F1($\%$)      \\ \hline
GMP & 94.31$\pm$2.21 & 90.98$\pm$2.15 & 95.85$\pm$1.84&92.97$\pm$2.02 &

 94.47$\pm$1.44 &  93.38$\pm$1.47 & 96.94$\pm$0.81 & 93.80$\pm$1.60& 84.79$\pm$1.53 & 86.14$\pm$2.13 & 94.36$\pm$0.48 & 85.15$\pm$1.55 \\

GAP & 95.06$\pm$1.14 &95.59$\pm$1.11 & 93.37$\pm$1.61 & 94.35$\pm$1.24 & 94.91$\pm$1.64 & 95.01$\pm$1.76 & 97.16$\pm$0.91 & 94.39$\pm$1.68 & 85.09$\pm$1.64 & 87.25$\pm$2.02 & 94.36$\pm$0.83& 85.60$\pm$1.81 \\
Attention Pooling & \textbf{96.94$\pm$1.10} &
  \textbf{97.83$\pm$1.53} &
  \textbf{94.93$\pm$1.26} &
  \textbf{96.23$\pm$1.02} &

  \textbf{95.85$\pm$1.27} &
  \textbf{95.74$\pm$1.20} &
  \textbf{97.73$\pm$0.45} &
  \textbf{95.54$\pm$1.59} &
  \textbf{87.12$\pm$1.55} &
  \textbf{87.20$\pm$1.64} &
  \textbf{95.20$\pm$0.64} &
  \textbf{86.98$\pm$1.27} \\ \Xhline{1.2pt}

\end{tabular}}
\end{table*}

\begin{table*}[]
\renewcommand{\arraystretch}{1.1}
\caption{Contribution of Attention Guided Augmentation}
\label{TAB:Retrain}
\centering
\resizebox{\textwidth}{!}{
\begin{tabular}{ccccccccccccc}
\Xhline{1.2pt}
\multicolumn{1}{c|}{\multirow{2}{*}{  \textbf{Augmentation}}} & \multicolumn{4}{c|}{\textbf{Dataset A}}             & \multicolumn{4}{c|}{\textbf{Dataset B}}            & \multicolumn{4}{c}{\textbf{Dataset C}} \\ \cline{2-13}
\multicolumn{1}{c|}{} &
  ACC($\%$)&
  SEN($\%$)&
 SPC($\%$)&
  \multicolumn{1}{c|}{F1($\%$)} &
  ACC($\%$)&
  SEN($\%$)&
 SPC($\%$)&
  \multicolumn{1}{c|}{F1($\%$)} &
  ACC($\%$)&
  SEN($\%$)&
 SPC($\%$)&
  F1($\%$)\\ \hline
\multicolumn{1}{c|}{$A^{M*}$} &
  94.69$\pm$1.38 &
  96.06$\pm$1.68 &
  92.82$\pm$1.50 &
  \multicolumn{1}{c|}{94.10$\pm$1.08} &
  93.56$\pm$1.51 &
  92.79$\pm$1.60&
  96.57$\pm$0.68 &
  \multicolumn{1}{c|}{92.65$\pm$1.43} &
  85.47$\pm$1.40&
  85.48$\pm$1.44 &
  94.64$\pm$0.54 &
  85.10$\pm$1.13 \\
\multicolumn{1}{c|}{$A^M + A^{D**}$} &
  95.81$\pm$1.23 &
  96.70$\pm$1.52 &
  94.86$\pm$2.23&
  \multicolumn{1}{c|}{95.59$\pm$1.34} &
  94.97$\pm$1.57 &
  95.10$\pm$1.34 &
  97.25$\pm$0.73 &
  \multicolumn{1}{c|}{94.84$\pm$1.38} &
  86.31$\pm$1.78 &
  87.12$\pm$1.82 &
  94.83$\pm$0.85&
  86.60$\pm$1.64 \\
\multicolumn{1}{c|}{$A^M + A^D + A^{P***}$} &
  \textbf{96.94$\pm$1.10} &
  \textbf{97.83$\pm$1.53} &
  \textbf{94.93$\pm$1.26} &
  \multicolumn{1}{c|}{\textbf{96.23$\pm$1.02}} &

  \textbf{95.85$\pm$1.27}&
  \textbf{95.74$\pm$1.20} &
   \textbf{97.73$\pm$0.45}&
  \multicolumn{1}{c|}{\textbf{95.54$\pm$1.59}} &
  \textbf{87.12$\pm$1.55} &
  \textbf{87.20$\pm$1.64} &
  \textbf{95.20$\pm$0.64} &
  \textbf{86.98$\pm$1.27} \\ \Xhline{1.2pt}
\multicolumn{13}{l}{*$A^M$: Attention Mix-up; **$A^D$: Attention Dimming; **$A^P$: Attention Patching.}
\end{tabular}}
\end{table*}

\begin{table*}[]
\renewcommand{\arraystretch}{1.1}
\caption{Comparison of L2 and Soft Distance Regularization}
\label{TAB:Regs}
\centering
\resizebox{\textwidth}{!}{
\begin{tabular}{c|cccc|cccc|cccc}
\Xhline{1.2pt}
\multirow{2}{*}{\textbf{Loss}} & \multicolumn{4}{c|}{\textbf{Dataset A}}          & \multicolumn{4}{c|}{\textbf{Dataset B}} & \multicolumn{4}{c}{\textbf{Dataset C}} \\ \cline{2-13}
   &ACC($\%$)           & SEN($\%$)   &SPC($\%$)   & F1($\%$)    & ACC($\%$)   & SEN($\%$)            &SPC($\%$)            & F1($\%$)    & ACC($\%$)    & SEN($\%$)   &SPC($\%$)   & F1($\%$)    \\ \hline
L2 & 95.93$\pm$0.61 & 96.36$\pm$0.85 & 95.48$\pm$0.71 & 95.83$\pm$0.79 & 94.62$\pm$0.86 & 93.75$\pm$0.91 & 97.05$\pm$0.68 & 93.75$\pm$0.93 & 83.99$\pm$0.94 & 85.33$\pm$1.34 & 94.11$\pm$0.69 & 84.80$\pm$0.83 \\
Soft Distance         & \textbf{96.94$\pm$1.10} &
  \textbf{97.83$\pm$1.53} &
  \textbf{94.93$\pm$1.26} &
  \multicolumn{1}{c|}{\textbf{96.23$\pm$1.02}} &
  \textbf{95.85$\pm$1.27}&
  \textbf{95.74$\pm$1.20} &
   \textbf{97.73$\pm$0.45}&
  \multicolumn{1}{c|}{\textbf{95.54$\pm$1.59}} &
  \textbf{87.12$\pm$1.55} &
  \textbf{87.20$\pm$1.64} &
  \textbf{95.20$\pm$0.64} &
  \textbf{86.98$\pm$1.27} \\ \Xhline{1.2pt}

\end{tabular}}
\end{table*}

\begin{figure*}[]
\centering
\includegraphics[width=181mm]{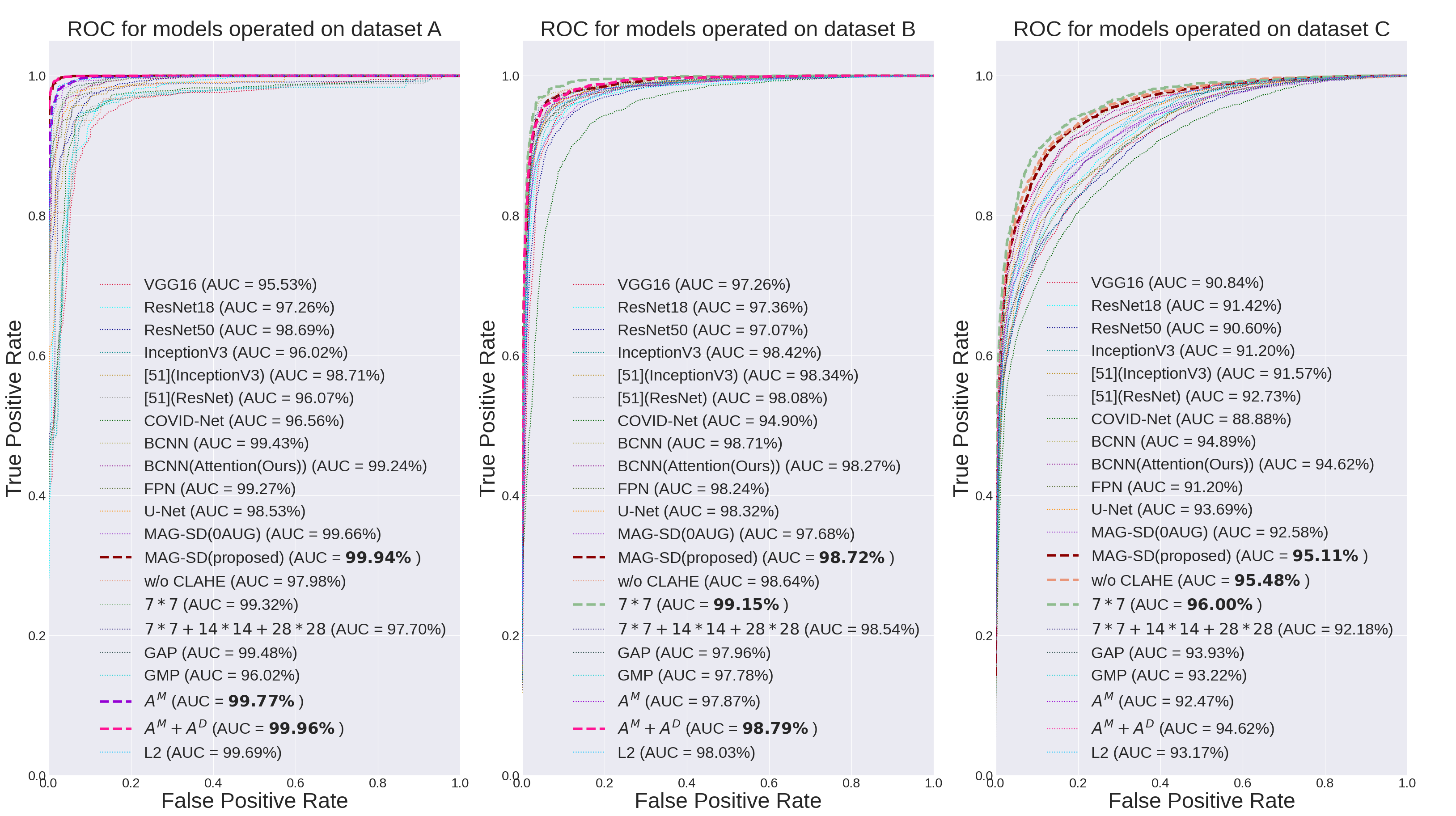}
\caption{Demonstration of ROC curves and AUC values. Three charts, from left to right, show the performance of all the trained models operating using datasets A,B and C respectively. Top-3 highest AUC values and their ROC curves are emphasized. Results demonstrate that comparing with baselines, results generated by our model has advantage in AUC value, which is over $0.5\%$ in dataset A, B and C. Architecture differences of our proposed method also influence the performance over datasets. Generally, MAG-SD(proposed) is the most stable model which stays in top-3 in all the datasets, which is a method given consideration to both generalizability and robustness.}
\label{Fig:ROC} 
\end{figure*}

\subsection{Evaluation Metrics}
Experiments were evaluated by several metrics. For Classification, Accuracy (ACC), Sensitivity (SEN), Specificity (SPC) and F1 score were employed. For multi-class datasets, mean value between classes were calculated to represent the final performance score of each model. Plots of receiver operating characteristic (ROC) curve and area under the curve (AUC) value were used to compare model functionality. Localization quality was quantified by intersection over union (IOU) which has been widely used in target detect and semantic segmentation task \cite{sedai2018deep}.
Accuracy describes the proportion of correctly classified targets, which is expressed in Eq. (\ref{eq13}).
\begin{equation}
\label{eq13}
    Accuracy = \frac{TP+TN}{TP+TN+FP+FN}
\end{equation}
where TP, TN, FP and FN stand for the number of true positive, true negative, false positive and false negative predictions.
Sensitivity, also known as true positive rate (TPR), is useful to measure the proportion of true positive predictions over all positive targets, which is defined in Eq. (\ref{eq14}).
\begin{equation}
\label{eq14}
    Sensitivity = \frac{TP}{TP+FN}
\end{equation}

Specificity, or true negative rate (TNR), is a ratio between the amount of true negative (TN) and false positive (FP) predictions, defined in Eq. (\ref{eq15}).
\begin{equation}
\label{eq15}
    Specificity = \frac{TN}{TN+FP}
\end{equation}
F1 Score considers the performance from both precision and recall, defined in Eq. (\ref{eq16}).
\begin{equation}
\label{eq16}
    F_1 = \frac{2TP}{2TP+FP+FN}
\end{equation}
IoU represents a value calculated by dividing the overlap of prediction and ground truth by their union. It could be defined straightforward in Eq. (\ref{eq17}), where $A_o$ and $A_u$ denote area of overlap and area of union respectively.
\begin{equation}
\label{eq17}
    IoU = \frac{A_o}{A_u}
\end{equation}

\subsection{Components Validation and Discussion}
\begin{figure*}[]
\centering
\includegraphics[width=180mm]{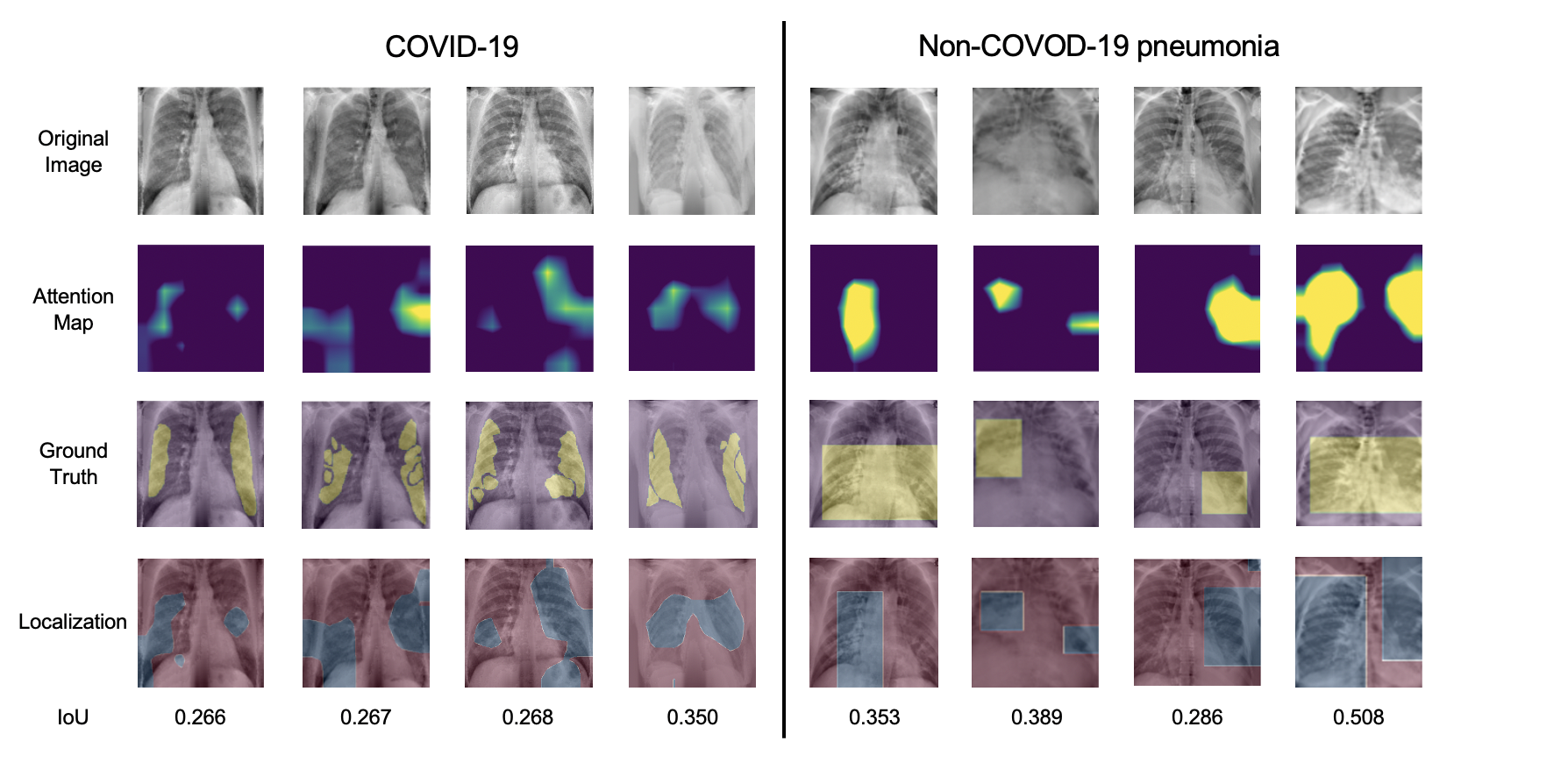}
\caption{Demonstration of pneumonia localization. Images were selected from \textit{Localization} dataset. COVID-19 cases has pixel-wise mask while bounding boxes were provided for other pneumonia. IoU was calculated for each prediction. Localization result was provided by apply threshold onto the attention map of each case. Results illustrated that attention focus on different area when detecting various classes.}
\label{FigAttenAna} 
\end{figure*}

The methods composed could be concluded into attention modules, attention guided data augmentation and soft distance regularization. Each component was studied by evaluating its improvement in classification performance, which has been quantified by metrics mentioned above. Performance gain was obtained by the following method: the proposed model was first trained on specific dataset with metrics, then, single component was changed or removed and reevaluate on the same dataset. For all the tested models, Mean value and standard deviation of ACC, SEN, SPC, F1 were recorded. Components validations were reported in Tab. \ref{TAB:CLAHE}, \ref{TAB:FeatureSizes}, \ref{TAB:Poolings}, \ref{TAB:Retrain} and \ref{TAB:Regs}. Inter-model comparisons could be found in Tab. \ref{TAB:Models} and Fig. \ref{Fig:ROC}. Regions interested the attention module were presented in Fig. \ref{FigAttenAna}. Parameters in all the experiments were maintained unchanged as possible for condition control. The model were trained on the same size of training set then evaluated on the same size of testing set.

\subsubsection{Architecture Comparing}
Advantages of architecture design has been deeply explored. It has been performed by evaluating classic coarse-grained deep neural networks (\textit{i.e. VGG16, ResNet18, ResNet50} and \textit{InceptionV3}), COVID-19 oriented architectures (\textit{i.e. \cite{narin2020automatic}(ResNet), \cite{narin2020automatic}(InceptionV3), COVID-Net-Large}), high performance fine-grained image classification structure (\textit{i.e. BCNN, BCNN(Attention)}) and multiscale feature fusion models (\textit{i.e. FPN, U-Net}).
Statistics analysis between these deep structures helped to explain our advantages in fine-grained feature extraction. It can be observed in Tab. \ref{TAB:Models} and Fig. \ref{Fig:ROC} that proposed model had noticeably better performance over others. For our model, accuracy on dataset A, B and C reached $96.94\%\pm1.10\%$, $95.85\%\pm1.27\%$, $87.12\%\pm1.55\%$ respectively and performance assessed by AUC are $99.94\%$, $98.72\%$, and $95.11\%$.

Comparing with classic models, our model was specialized for COVID-19 image classification and attention guided training phase had advantage in fine-grained visual classification task.
Most of the other COVID-19 oriented models  presented better performance than classic models, however, none of them applied attention mechanism or considered fine-grained features, which impacted their accuracy on large scale, multi-class dataset such as Dataset B and C.
Comparison between FPN, U-Net and classic models demonstrated that FPN presented results over InceptionV3 in Dataset A and B.
In Dataset C, U-Net had higher accuracy than FPN, which exceeded ResNet50. Results indicated that multiscale feature fusion models reached competitive results using relatively simple structures comparing with classic deep models, which left us a hint that multiscale attention might be a possible route to improve.

BCNN was a FGVC model with stable performance on multiple datasets. In order to evaluate the generalization ability of our attention module, multiscale attention and attention pooling were transported to BCNN to train BCNN(Attention). Statistically, BCNN reached $96.00\%\pm1.52\%$, $94.41\%\pm1.37\%$, $84.36\%\pm1.84\%$ in Accuracy, which was competitive in all the evaluated models. Attention modules remarkably boosted the performance of BCNN, exceeded our proposed method in Dataset A (SPC) and Dataset B (SEN), which were $96.16\%\pm1.74\%$, $96.61\%\pm2.00\%$ respectively.

\begin{figure*}[]
\centering
\includegraphics[width=180mm]{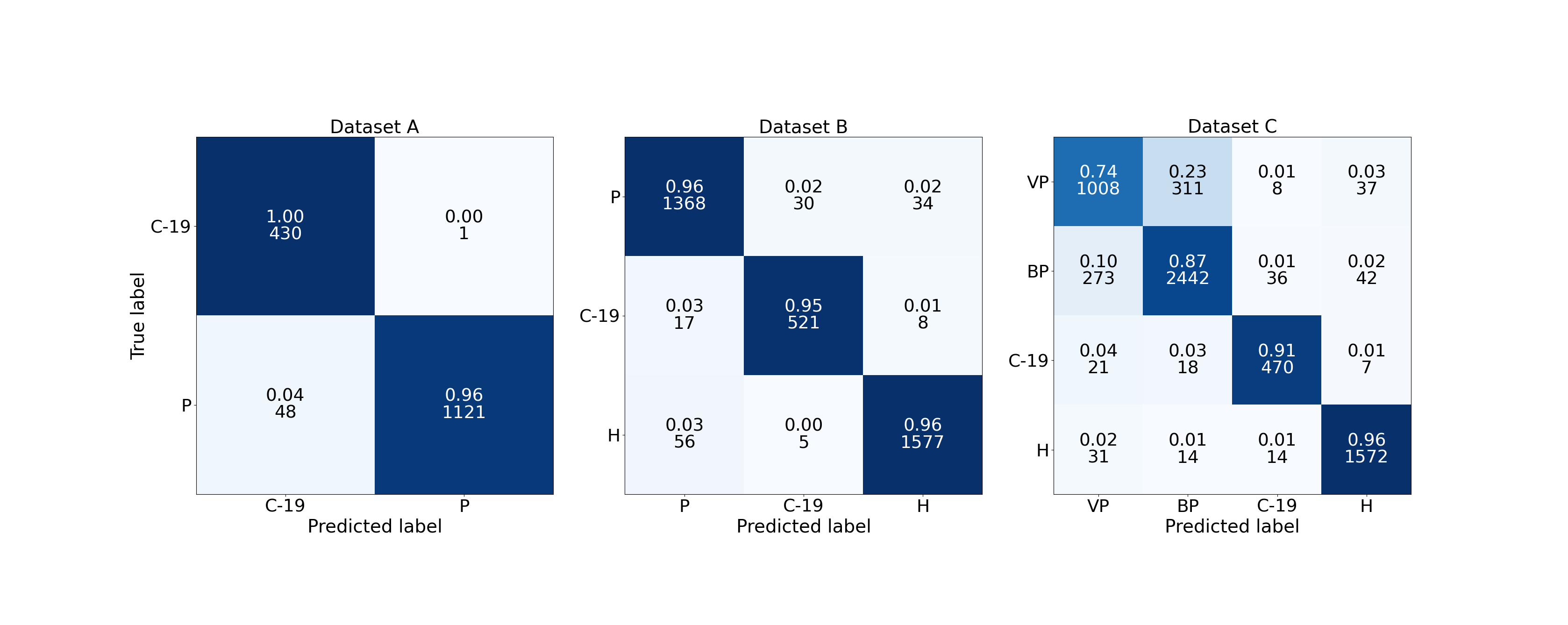}
\caption{Three charts of confusion matrices generated by proposed MAG-SD, demonstrating the distribution of predictions. The color of confusion matrices depend on the normalized values of predictions for a better visualization, which are placed at the top of each grid. The number of predictions are placed below the normalized values. Symbols used in figure are denoted as: P: Non-COVID19 Pneumonia, VP: Viral Pneumonia, BP: Bacterial Pneumonia, C-19: COVID-19, and H: Healthy. }
\label{Fig:CM} 
\end{figure*}

\subsubsection{CLAHE Preprocessing}
Images collected by different devices were probably distinct in contrast due to configuration variety. CLAHE was employed to relieve the noise brought by contrast distribution. Tab. \ref{TAB:CLAHE} showed the result that CLAHE obviously improved the performance of proposed model and raised over $1.5\%$ Accuracy on average. Model trained without CLAHE had notable higher standard deviation value. Larger datasets such as Dataset B and C were reported to have more performance gain.

\subsubsection{Multiscale Attention Generator and Attention Pooling}
Normally, state-of-the-art coarse-grained CNN models suffer from similar global features when dealing FGVC. Under this circumstance, models have to depend on local features, which could be effectively localized by our multiscale attention module. Models trained with attention module (\textit{i.e. MAG-SD(0AUG)}) and baseline model (\textit{i.e. ResNet50}) were compared in Tab.\ref{TAB:Models}, and Fig. \ref{Fig:ROC}. Results revealed that proposed model surpasses the baseline on dataset A, B and C using most of the benchmarks. In dataset B (SPC), ResNet50 has slight advantage. Comparing with AUC, MAG-SD(0AUG) was $2\%$ over ResNet50.
Furthermore, two parts of attention module, attention generating and attention pooling has been investigated separately. Firstly, models were compiled to assess multiscale attention, with $1,2$ or $3$ size of attention maps considered. Results presented in Tab. \ref{TAB:FeatureSizes} and Fig. \ref{Fig:ROC} showed the model considering two feature maps achieved the best performance in all three datasets. Possible explanation was that the proposed attention module was too simple to locate valuable fine grained feature on low-level feature maps. Instead of importing meaningful location information, noise was brought into the proposed model.
Secondly, we evaluated attention pooling module with models trained with other commonly used pooling methods such as global average pooling (GAP) or global max pooling (GMP). Results on pooling methods were presented in Tab. \ref{TAB:Poolings} and Fig. \ref{Fig:ROC}, demonstrating that attention pooling surpassed GAP and GMP in all three datasets.

\subsubsection{Attention Guided Augmentation}
The generated attention maps emphasized local feature that interested the model, which could be used to effectively guide data augmentation in Fig. \ref{FigReT}. Models trained with $0,1,2$ or $3$ augmentation were discussed in experiment. In the case of $1$ augmentation, attention mixup was selected. $2$ augmentations model included attention mixup and attention patching. The results were presented in Tab. \ref{TAB:Retrain} and Fig. \ref{Fig:ROC}. The table reflects that model with all three augmentations had better performance, however, AUC value showed that in dataset A and B, two augmentations was advantageous. The proposed augmentations emphasized data according to attention map, minimizing negative effect caused by random augmentations.

\subsubsection{Soft Distance Regularization}
Soft distance regularization was presented to relieve augmentation variance. Experiments have been composed to compare it with $L2$ distance regularization. Tab. \ref{TAB:Regs} and Fig. \ref{Fig:ROC} illustrated that it surpassed $L2$ in mean value, but, inferior in standard deviation. Constraint between auxiliary vector and primary vector screen the false prediction introduced by attention guided augmentations. Regularization was calculated between ground truth and auxiliary vector when primary vector cannot provide reliable prediction, keeping the final result away from local minima. $L2$ compared all predictions directly with ground truth, which has higher stability and sidestepped the disturbances introduced by primary prediction.

\subsubsection{Attention Based Infection Localization}
Technically, attention improved the models by roughly localize the part with high activation intensity. This characteristic of attention inspired us to try \textit{MAG-SD} on localization topics. The models were trained on the \textit{Dataset B} we proposed, then test on \textit{Localization} dataset demonstrated in Fig. \ref{FigAttenAna}. It included COVID-19 cases with pixel-wise segmentation and non-COVID-19 cases with bounding box for pneumonia infection. Attention maps $A$ were upsampled from $7*7$ to $224*224$. Localization masks for COVID-19 cases were extracted by applying threshold to the attention maps. Bounding boxes for other pneumonia were produced by simply enclosing the localization masks with rectangles. IoU was calculated to evaluate the quality of localization. Image showed that the attention module we proposed could roughly indicate the position of different type of pneumonia with over 0.25 IoU score. Attention map emphasized the influential part from the input image effectively.

\subsection{Distribution Analysis}
As we imported multiple fine-grained classes into this topic, it was necessary to report the distribution of our prediction result, which has been shown using confusion matrix in Fig. \ref{Fig:CM}. MAG-SD has been selected to generate the charts to represent the classification result of deep learning models. It could be inferred that the model was suitable for searching definitive features from cases showed in dataset A and B as most of the cases were located on the diagonal line of matrices. In dataset C, classification result between viral pneumonia and bacterial pneumonia was significantly inferior than others, which impacted the global performance of the classification model. These results proved the arguments reviewed in Section. \ref{SEC:PuXR}, indicated that the CXR visual appearance between viral and bacterial pneumonia was insufficient to make accurate diagnosis.

\section{Conclusions}
\label{SEC:CONCLUSIONs}

We have presented \textit{MAG-SD} for automatic COVID-19 CXR image classification that reached the state-of-the-art on our dataset. The proposed novel method treated this topic as a fine-grained image classification task, utilizing local features efficiently under the guidance of attention mechanism. Attention maps were generated using multiscale features then used as a reference to data augmentation, helping the model to overcome the lack of COVID-19 cases.
The proposed network learned to weight the predictions from both primary and auxiliary training pathways by calculating soft distances between vectors, gaining improvements by screening noise generated by augmentations.

Findings of our exploration were demonstrated and discussed in Section. \ref{SEC:EXPERIMENT}. The results indicated the great potential of applying advanced pattern recognition model to clinical diagnosis and epidemic screening. Trained on the clinical knowledge acquired by physicians, our model was capable to extract fine-grained spatial features for COVID-19. Attention was applied in both feature extraction and augmentation stage, which helped to localize pneumonia infection and accrete the data effectively as part of weakly supervised method. Attention module also shows its capability in different models. It could be interesting to design more auxiliary training strategies to guide the model to an optimal solution. Positive feedback on soft distance regularization proved that our method considered auxiliary predictions and eliminated label noise simultaneously, however, hard threshold may limit its adaptability in complicated data.

Although deep learning methods seem promising in clinical diagnosis and pandemic screening, lacking of prior knowledge is always the Achilles' Heel. Supervised learning method, such as \textit{MAG-SD} we proposed, have to be trained on expensive labeled data. Newly occurred or rare diseases without available data may not be classified properly. Abnormal detecting and clustering model could be proposed as a guidance for supervised models to alleviate the limitations, which is part of our future work.

\section*{Acknowledgments}
This work was supported by the National Natural Science Foundation of China (Grant No.KYZ043718114) and Biomedical Engineering Interdisciplinary Research Fund of Shanghai Jiao Tong University (Grant No.YG2020YQ17). The authors would like to thank the institutes who generously open-sourced image database.

\bibliographystyle{unsrt}
\bibliography{arxiv.bib}

\end{document}